\documentclass[12pt]{article}
\usepackage{graphicx}
\def\beq{\begin{equation}}
\def\eeq{\end{equation}}
\def\bea{\begin{eqnarray}}
\def\eea{\end{eqnarray}}

\begin{document}
\setcounter{page}{1}
\setcounter{equation}{0}
\baselineskip22pt
\parskip18pt
\topmargin=-12pt

\renewcommand{\thefootnote}{*}
\begin{center}
{\Large\bf Deghosting of Ocean Bottom Cable Data :Two approaches}  \\[1.5cm] {\bf Jagmeet 
Singh}\footnote{E-mail: jagmeetus@yahoo.com} and {\bf K. Subramanyam}\\ 
{\it Geodata Processing and Interpretation Centre, KDMIPE Campus, ONGC Ltd., Dehradun - 248 195, India} \\
\end{center} 

\vskip1.0cm

\begin{abstract}
Two filter based approaches for deghosting of Ocean Bottom Cable data are presented. One of them is phase shifting of geophone followed by least square matching of cross ghosted geophone and hydrophone data.In the second approach,phase shifting of geophone is followed by direct matching of geophone's amplitude spectrum with that of the hydrophone within the seismic bandwidth.Results obtained from both approaches were found to be at par with one of the proprietary softwares available with ONGC.
\end{abstract} 

\vskip1.0cm

\newpage

\baselineskip22pt
\parskip18pt

\section{Introduction}
Towed streamer operations in congested areas lead to gaps in 3-D coverage and hence the  OBC(Ocean Bottom Cable) method employing detectors on the ocean bottom becomes a necessity.In order to ensure that high quality seismic data is obtained using this method one has to address the problem of water column reverberation at the acquisition stage itself.Every primary arrival at a detector location is followed by secondary arrivals due to reverberation of the seismic energy between the water surface and  water bottom.If the water depth is small(i.e. below 10m),the time gap between the primary and secondary arrivals is  small and  the undesired secondaries can be removed by deconvolution algorithms.

However if the water depth is high,deconvolution can not remove the multiples because the time gaps are large.The solution is to record both geophone and hydrophone data at every receiver location.Since the geophone records velocity whereas the hydrophone records pressure,the sum of the two is devoid of the  downgoing part of the wavefield which is recorded with opposite polarities on the hydrophone and geophone.In frequencydomain terms there are notches in the frequency spectrum of hydrophone data, at $f=n f_0$ ,where $f_0 = v/2z$($v$ is the acoustic  velocity in water and $z$ is the receiver depth),which fall within the seismic  bandwidth.The notches in the geophone spectrum occur at $f=(n+1/2)f_0$ so that the  sum of the two spectra is free of notches.   

Most of the earlier methods aimed at deghosting were scalar methods.A scalar  was designed to be multiplied with the geophone data in such a way that the  autocorrelogram of the sum of hydrophone and geophone data was as spiky as  possible(Barr and Sanders(1989)\cite{Barr},Dragoset and Barr(1994)\cite{Dragoset},Barr(1997)\cite{Barr1}).Clearly  these methods were inadequate,in theory,as well as in practical results obtained  from such methods.Hydrophone and geophone response are different,geophone  ground coupling  varies from place to place due to which it is clear that a filter  should be designed(at every receiver location) and applied to either of the two (preferably geophone) before the two are summed.
 
One such filter based technique has been suggested by Robert Soubaras(1996)\cite{Robert}. In this paper we have used this approach with some modifications and tested the  same on synthetic and real data,as well as a frequency domain approach.The results  obtained from these methods were compared with the results obtained from  proprietary software available with ONGC.The paper is organized as follows.In  section 2 we describe the method  followed by results obtained by the two methods  in section 3.

\section{Designing the filter}

Figure 1 shows an incident wavefield I(z) just above the water bottom,where the delay z corresponds to the sampling interval.If U and D denote upgoing and downgoing wavefields just above the water bottom,then the hydrophone and geophone record pressure H and velocity G respectively given as
\beq
H = U + D 
\eeq
\beq
G= \frac{U - D}{I_0},
\eeq

where $I_0$  is the acoustic impedance of water.The constant $I_0$ will be dropped from further calculations.From the formulae (1) above we see that if D=0,H and G seem to be in phase.Clearly pressure and velocity have to have a phase difference of 90 degrees,so we have to bear in mind,at the outset,that phase shifting is required. If Z denotes the delay corresponding to two way time of travel in the water column then U(z) and D(z) are given as follows:-
\beq
U(z)= I(z) (1- RZ + R^2Z^2 +.......)
    = \frac {I(z)}{1+RZ}                                       
\eeq
\beq
D(z)=I(z)(-Z + RZ^2 - R^2Z^3 +.....)
    =\frac{-I(z) Z}{1+RZ},
\eeq

where R is the reflection coefficient of the water bottom.Using the above 4 equations it follows that
\beq
H(z)= \frac{(1-Z)I(z)}{1 + RZ}
\eeq
\beq
G(z)= \frac{(1+Z)I(z)}{1 + RZ}
\eeq

From (5) and (6) it follows that
\beq
(1+Z)H(z) = (1-Z)G(z)                                                               
\eeq

For the reason of phase shift mentioned above and also because the hydrophone and geophone response are never the same, it would be more appropriate to write Eqn.(7) as
\beq
(1+Z)H(z) = (1-Z)G(z)PF(z),
\eeq
where P is a phase shifting(90 degrees) operation and F(z) is a filter that accounts for the difference in  impulse response of the two phones.So our scheme of operation is the following.We introduce the necessary phase shift and the convolutions.The filter F is then designed in the time domain in such a way that the R.H.S. of (8) matches L.H.S. of (8) in the least square sense i.e. a Wiener filter is designed.

The phase shifted geophone is then convolved with the filter,multiplied by a suitable scalar and added to the hydrophone.We have to keep in mind the following.In using delay Z(corresponding to v/2z),we have assumed vertical or near vertical bouncing of rays in the water column.With increasing offset,we must go deeper i.e. take design windows deeper(in designing F(z) as per (8)) so that our assumption is satisfied.Our experiments have confirmed that separating the filtering action into P and F leads to better results as it puts less burden on the Wiener filter F.

The scalar required follows from (5) and (6).We see that
\beq
H(z) + \frac{1+R}{1-R} G(z) = \frac{2}{1-R}I(z)
\eeq

From the above equation we see that the required scalar (1+R)/(1-R) accomplishes the rest of the job once filtering is done.R is typically 0.4.We have, however, used amplitude equalisation of the two phones over a moving window as a second step once filtering is applied.

\section{Results and Discussion}

Figure 2 shows synthetic seismograms for geophone data for a small water depth.Figure 3 shows synthetic hydrophone data for the same water depth.Figure 4 shows the sum of the hydrophone and filtered geophone which is free of the ghost.Figure 5 shows the deghosted and stacked output for a real dataset using Wiener filter approach.The same compares well with the output shown in Figure 7 obtained from a proprietary software of ONGC using identical velocities.In the frequency domain approach apart from phase shifting, the amplitude spectrum of the geophone is matched with that of the hydrophone in the seismic bandwidth.The stack obtained from this method is shown in Figure 6.Figure 8 shows the autocorrelation averaged over a range of CDP values obtained from the Wiener filter(above part of figure), and also the autocorrelation averaged over the same range obtained from the proprietary software mentioned above.Figure 9 shows the averaged autocorrelation obtained from the frequency domain approach. Autocorrelations obtained from our methods are sharper.

As stated above,reverberations in the water column are assumed to follow near vertical ray paths for ensuring which we take our filter design windows deeper with increasing offset.A more appropriate method would be to reject 1-d approximation altogether and adapt our scheme for angled reverberations.The results of such a scheme will be presented in a forthcoming paper.

\section{Acknowledgements}
We express our gratitude to Director(Exploration),ONGC Ltd.for his kind permission to publish this paper.We are indebted to Dr. C.H. Mehta,Head Geopic,ONGC Ltd.for many useful  discussions and suggestions on the subject.We are very grateful to Shri Kunal Niyogi,Head Processing,Geopic,ONGC Ltd.for his encouragement and support throughout this work.

\vskip1.0cm

\newpage

\begin{figure}[htbp]
\begin{center}
\includegraphics[height=8cm,width=14cm,angle=0]{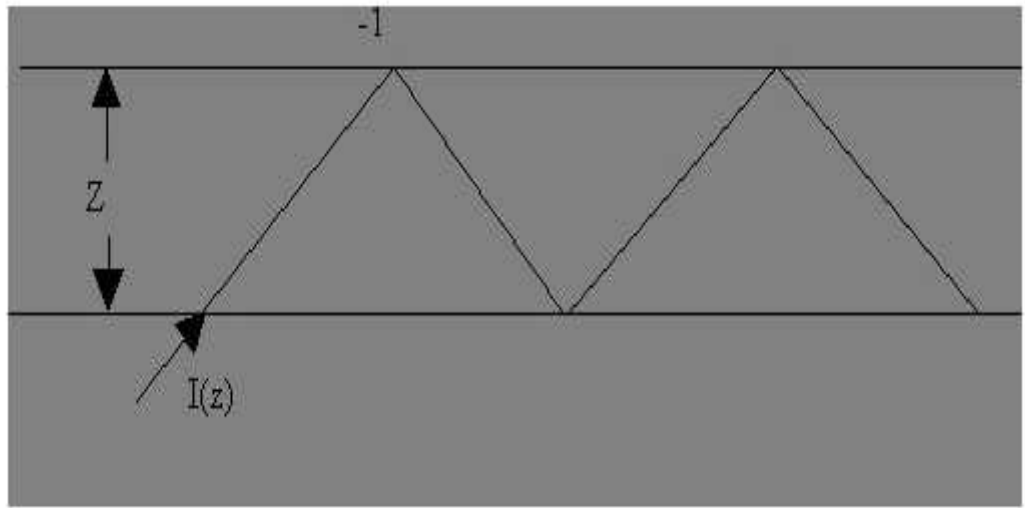}
\caption{\small The incident wavefield  is shown to bounce back and forth between
the water surface and the water bottom.}
\label{1}
\end{center}
\end{figure}

\begin{figure}[htbp]
\begin{center}
\includegraphics[height=8cm,width=6cm,angle=0]{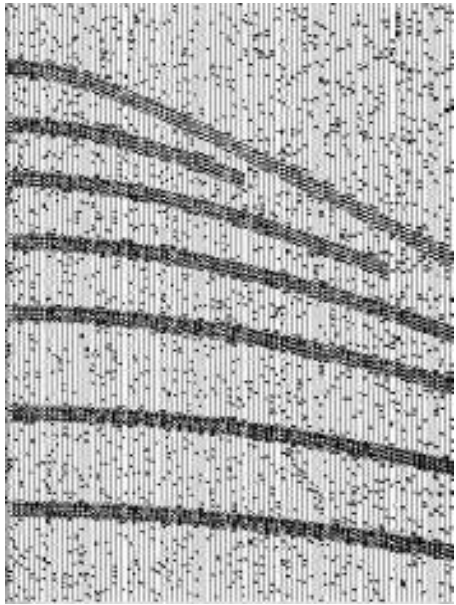}
\caption{\small Synthetic geophone gather.}
\label{2}
\end{center}
\end{figure}

\begin{figure}[htbp]
\begin{center}
\includegraphics[height=8cm,width=6cm,angle=0]{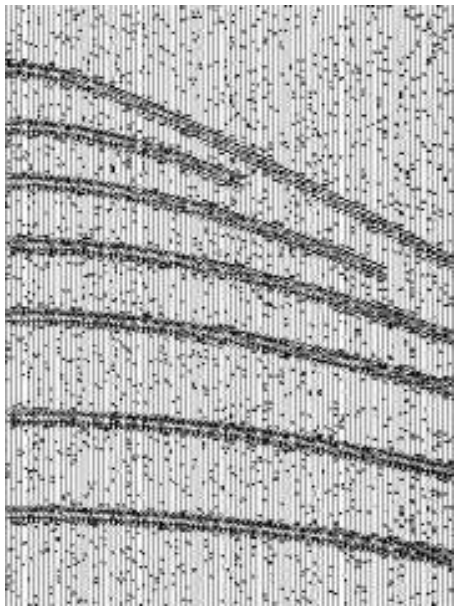}
\caption{\small Synthetic hydrophone gather.}
\label{3}
\end{center}
\end{figure}

\begin{figure}[htbp]
\begin{center}
\includegraphics[height=8cm,width=6cm,angle=0]{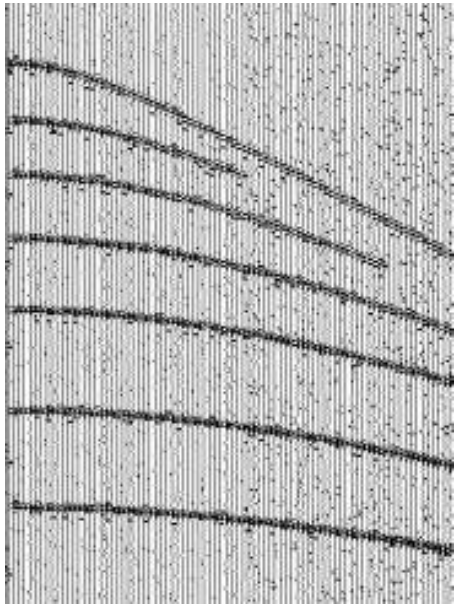}
\caption{\small Output obtained after filtering the geophone and adding to the hydrophone.}
\label{4}
\end{center}
\end{figure}

\begin{figure}
\begin{center}
\includegraphics[height=13cm,width=10cm,angle=0]{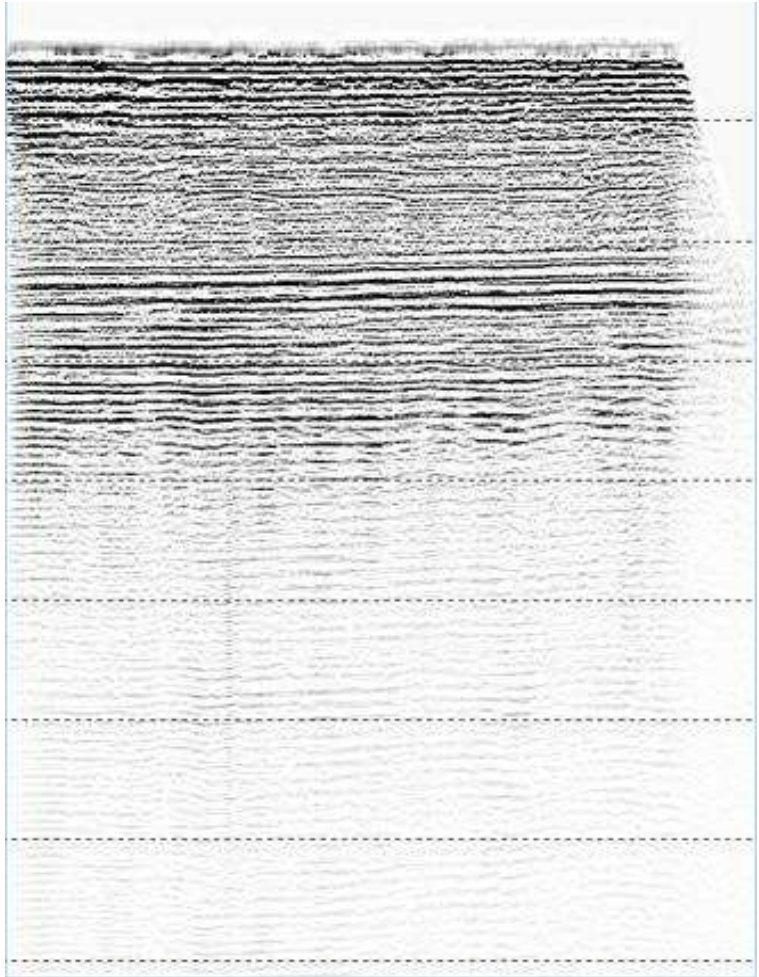}
\caption{\small Stacked output obtained from the Wiener filter approach.}
\label{5}
\end{center}
\end{figure}

\begin{figure}
\begin{center}
\includegraphics[height=13cm,width=10cm,angle=0]{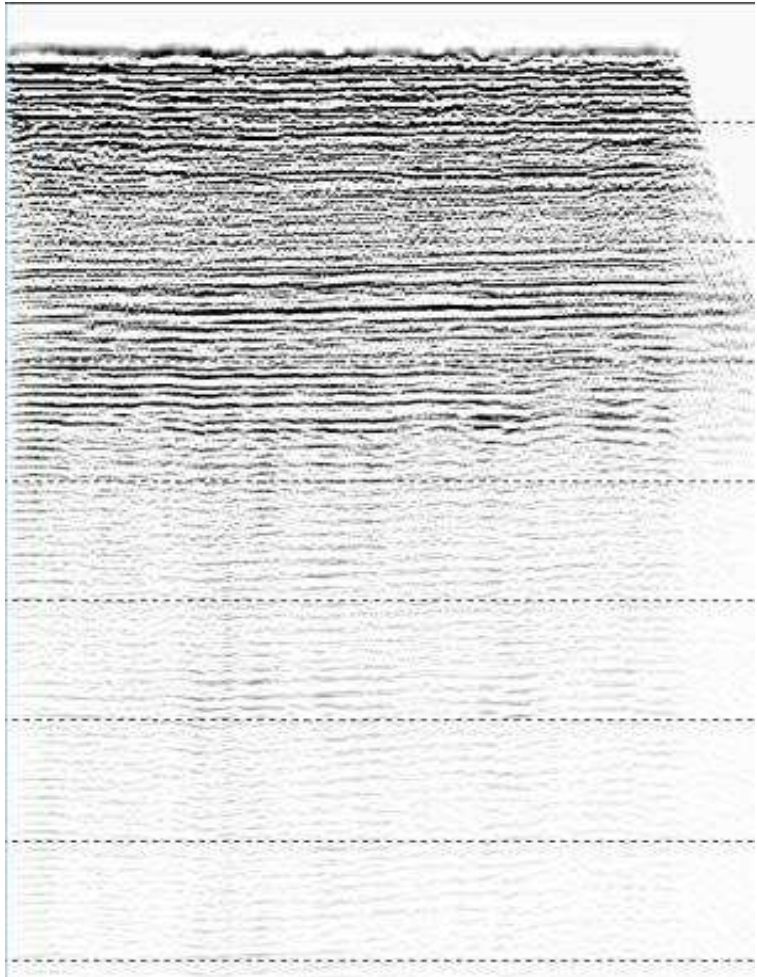}
\caption{\small Stacked output obtained from the frequency domain approach.}
\label{6}
\end{center}
\end{figure}

\begin{figure}[htbp]
\begin{center}
\includegraphics[height=13cm,width=10cm,angle=0]{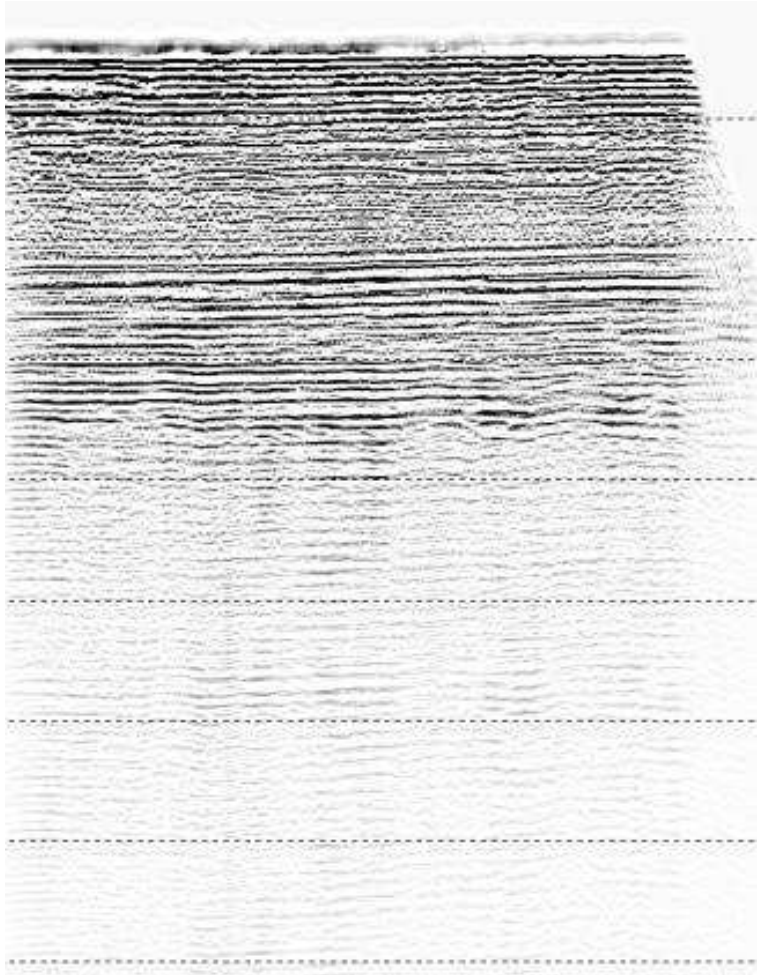}
\caption{\small Stacked output obtained from the proprietary software.}
\label{7}
\end{center}
\end{figure}

\begin{figure}[htbp]
\begin{center}
\includegraphics[height=10cm,width=10cm,angle=0]{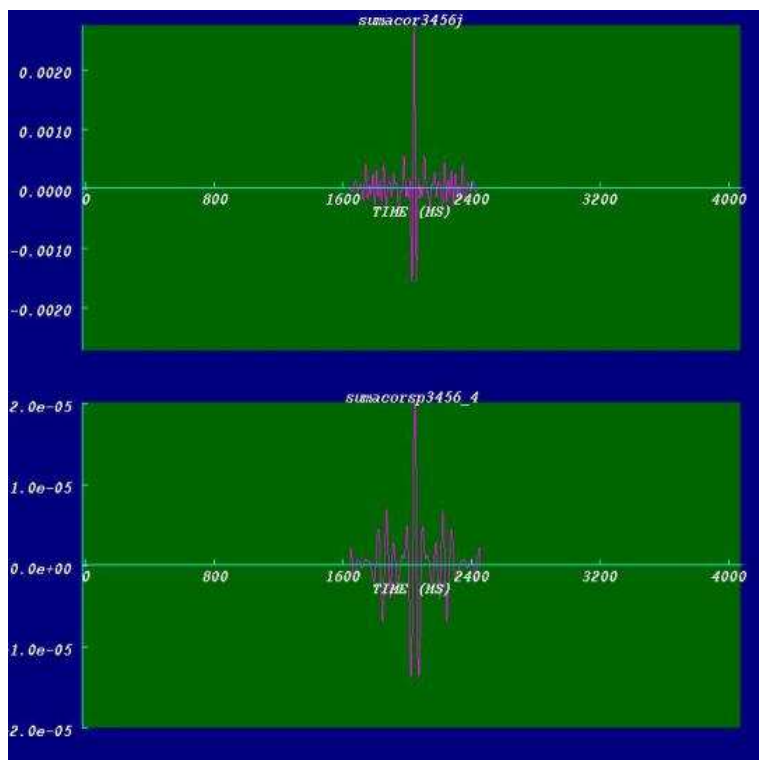}
\caption{\small The upper part of the figure shows the averaged autocorrelation obtained from the Wiener filter approach whereas the part below shows the same obtained from the proprietary software.}
\label{8}
\end{center}
\end{figure}

\begin{figure}[htbp]
\begin{center}
\includegraphics[height=4cm,width=10cm,angle=0]{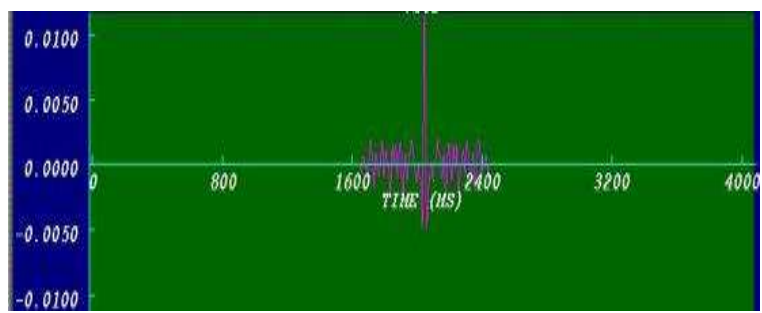}
\caption{\small Averaged autocorrelation obtained from the frequency domain approach.}
\label{9}
\end{center}
\end{figure}

\end{document}